\documentclass[aps,pra,twocolumn,superscriptaddress,nofootinbib,longbibliography]{revtex4-2}

\usepackage{amsmath,amssymb,bm,mathtools}
\usepackage{graphicx}
\usepackage{xcolor}
\usepackage[colorlinks=true,linkcolor=blue,citecolor=blue,urlcolor=blue]{hyperref}
\usepackage{microtype}
\newcommand{\ii}{\mathrm{i}}

\newcommand{\dd}{\mathrm{d}}
\newcommand{\GBZ}{\mathrm{GBZ}}

\newcommand{\UT}{U_T}
\newcommand{\UtwoT}{U_{2T}}
\newcommand{\corner}{\mathrm{corner}}

\newcommand{\eff}{\mathrm{eff}}
\newcommand{\intm}{\mathrm{int}}
\newcommand{\LD}{\mathrm{LD}}
\newcommand{\LU}{\mathrm{LU}}
\newcommand{\RD}{\mathrm{RD}}
\newcommand{\RU}{\mathrm{RU}}

\newcommand{\calC}{\mathcal C}
\newcommand{\calA}{\mathcal A}

\newcommand{\calN}{\mathcal N}
\newcommand{\maybegraphics}[2]{%
\IfFileExists{#2}{\includegraphics[width=#1]{#2}}{%
\fbox{\begin{minipage}[c][0.18\textheight][c]{#1}\centering missing figure file\\\texttt{\detokenize{#2}}\end{minipage}}}}

\begin{document}

\title{Complex-gauge control of anomalous Floquet corner responses in a non-Hermitian physical-synthetic photonic lattice}

\author{W. C. Ning}
\affiliation{College of Physics and Materials Science, Tianjin Normal University, Tianjin 300387, China}

\author{X. Z. Zhang}
\email{zhangxz@tjnu.edu.cn}
\affiliation{College of Physics and Materials Science, Tianjin Normal University, Tianjin 300387, China}

\date{\today}

\begin{abstract}
We propose a non-Hermitian Floquet photonic lattice formed by a physical resonator coordinate and a synthetic frequency coordinate.  A two-step modulation protocol realizes a chiral walk in this physical-synthetic plane, with a real synthetic flux controlling loop interference and imaginary gauge fields controlling non-reciprocal envelopes.  We show that anomalous corner pairs at quasienergies zero and \(\pi/T\) exhibit three distinct layers of physics.  A non-Bloch higher-order construction predicts whether the \(0/\pi\) corner pair exists under open boundaries.  The imaginary gauge fields select where the right eigenmodes accumulate.  The real flux controls the local interference matrix element that determines whether the doubled-period optical response is visible.  As a result, the same topological coexistence sector can be bright, skin-dark, or flux-dark in a local optical measurement.  We further show that the complex gauge can tune an exceptional point of the two-period corner propagator.  At this point the anomalous response keeps its doubled-period sign alternation, but its envelope becomes algebraic because of a Jordan block.  These results provide a photonic route to separate topological existence, skin-selected localization, optical visibility, and defective two-period dynamics in a non-Hermitian synthetic dimension.
\end{abstract}

\maketitle

\section{Introduction}

Dynamically modulated photonic systems provide a direct way to build lattices in mode space.  In a resonator array, the resonator index is a physical coordinate, while a ladder of frequency modes acts as a synthetic coordinate.  Resonant frequency conversion generates hopping along the synthetic direction, and modulation phases act as Peierls phases.  These properties make photonic synthetic dimensions a natural setting for Floquet walks, artificial gauge fields, and topological boundary responses \cite{Yuan2018,Ozawa2019,Dutt2019,Dutt2020,Lustig2019,Dutt2020HOTI}.

Floquet systems have a feature that is absent in static band structures.  A chiral Floquet walk has two special quasienergy gaps, one at zero and one at \(\pi/T\), and the two gaps can carry independent topological indices.  If a boundary supports both a zero mode and a \(\pi\) mode, their one-period multipliers differ by a minus sign.  A coherent local superposition then produces a response that alternates from period to period.  We refer to this subharmonic interference signal as the doubled-period optical response, or equivalently as the \(2T\) signal after this definition \cite{Asboth2013,Rudner2013}.  This response is not a static boundary intensity.  It is an interference effect between two anomalous Floquet sectors \cite{Kitagawa2010,Kitagawa2012,Asboth2013,Rudner2013,Lindner2011,Rechtsman2013,Maczewsky2017,Mukherjee2017}.

Non-Hermitian photonics introduces a second layer of control \cite{Bender1998,ElGanainy2018,Miri2019,Ashida2020,Bergholtz2021}.  Direction-dependent frequency conversion, asymmetric attenuation, and gain/loss engineering can implement imaginary gauge fields.  Under open boundaries these fields generate non-Hermitian skin accumulation, so the ordinary Bloch momentum no longer gives the correct bulk description.  It must be replaced by a generalized Brillouin zone, and topological indices have to be evaluated in non-Bloch variables \cite{Hatano1996,Kunst2018,Yao2018,Yao2018Chern,Yokomizo2019,Okuma2020,Kawabata2019}.  This distinction is especially important for higher-order boundary physics \cite{Benalcazar2017Science,Benalcazar2017PRB,Schindler2018}.  Skin accumulation can select a corner for the right eigenmodes without changing the existence of the topological corner states \cite{Lee2019,Kawabata2020}.

The present work asks a concrete optical question.  If a non-Bloch higher-order construction predicts a \(0/\pi\) corner pair, does a local detector necessarily observe a strong doubled-period response?  We show that the answer is no.  The topology determines whether the corner pair exists.  The imaginary gauge fields determine whether the two right eigenmodes are co-localized in the same corner window.  The real synthetic flux determines whether the local interference matrix element is constructive or destructive.  Thus one topological coexistence sector can appear in three different ways in an optical measurement: bright, skin-dark, or flux-dark.  This is the main distinction between the present proposal and a conventional observation of corner localization.  The measured object is the subharmonic interference response of an anomalous \(0/\pi\) pair, not only the intensity profile of a single corner state.

The same complex-gauge structure also controls the diagonalizability of the corner dynamics.  A zero mode and a \(\pi\) mode cannot coalesce as eigenvectors of the one-period Floquet operator while keeping their quasienergy labels, since their multipliers are \(+1\) and \(-1\).  They can, however, become defective in the two-period corner propagator, where both sectors are folded to the same multiplier.  At this exceptional point, the doubled-period optical response is not created from nothing.  Rather, an existing \(0/\pi\) response acquires a Jordan-block algebraic envelope, producing a defective doubled-period response.

The remainder of this paper is organized as follows.  Section~II introduces the photonic physical-synthetic lattice and the two-step complex-gauge drive.  Section~III derives the \(0/\pi\) corner subspace and the balanced interferometric readout of the local doubled-period response.  Section~IV formulates the edge non-Bloch construction for the higher-order corner diagnostic.  Section~V separates topological existence from optical visibility and identifies the bright, skin-dark, and flux-dark regimes.  Section~VI develops the exceptional-point theory of the two-period corner propagator.  Section~VII summarizes the optical diagnostics and the experimental observables.

\section{Photonic physical-synthetic Floquet lattice}

We consider a one-dimensional array of optical resonators indexed by \(x=1,\ldots,L_x\).  Each resonator supports a ladder of frequency modes
\begin{equation}
    \omega_w=\omega_0+w\Omega,
    \qquad
    w=1,\ldots,L_w .
    \label{freq_ladder}
\end{equation}
The frequency index \(w\) is treated as a synthetic coordinate.  Each physical-synthetic site \((x,w)\) contains two internal optical modes, denoted by \(A\) and \(B\).  These two modes may represent two paths, two polarizations, or two resonator supermodes.  They are the physical carriers of the \(\sigma_z\) drift and the \(\sigma_x\) mixing operations introduced below.  A single-excitation field, or equivalently a classical slowly varying optical envelope, is written as
\begin{equation}
    |\Psi\rangle
    =
    \sum_{x,w}|x,w\rangle
    \otimes
    \begin{pmatrix}
        a_{x,w,A}\\
        a_{x,w,B}
    \end{pmatrix} .
    \label{state}
\end{equation}
The tight-binding sites in Eq.~\eqref{state} are optical modes.  A hopping matrix element represents resonant conversion between two modes, not the motion of a material particle in real space.

The evolution over one period is generated by two modulation windows,
\begin{equation}
    \UT
    =
    \exp\left(-\ii H_2T/2\right)
    \exp\left(-\ii H_1T/2\right).
    \label{UT}
\end{equation}
The first window implements a non-reciprocal synthetic-frequency drift,
\begin{equation}
    H_1
    =
    \frac{t_w}{2\ii}
    \left(
        e^{\gamma_w+\ii\phi_w}T_w
        -
        e^{-\gamma_w-\ii\phi_w}T_w^{-1}
    \right)
    \otimes\sigma_z .
    \label{H1}
\end{equation}
Here \(T_w\) translates \(w\to w+1\).  The real phase \(\phi_w\) is a Peierls phase along the synthetic direction, while \(\gamma_w\) makes upward and downward frequency conversion unequal.  Because this step is proportional to \(\sigma_z\), the two internal modes acquire opposite synthetic drifts.

The second window implements flux-threaded internal mixing along the physical direction,
\begin{equation}
    H_2
    =
    \left[
        M
        +
        t_x
        \left(
            e^{\gamma_x}T_x(\Phi)
            +
            e^{-\gamma_x}T_x^{-1}(\Phi)
        \right)
    \right]
    \otimes\sigma_x .
    \label{H2}
\end{equation}
The Landau-gauge translation is
\begin{equation}
    T_x(\Phi)|x,w\rangle
    =
    e^{\ii\Phi w}|x+1,w\rangle .
    \label{TxPhi}
\end{equation}
When a mode moves around an elementary physical-synthetic plaquette, the accumulated phase is set by \(\Phi\).  Thus \(\Phi\) is a real synthetic magnetic flux.  The factor \(e^{\gamma_x}\) is the corresponding imaginary gauge factor along the physical direction.  These two types of gauge control play different roles throughout the paper: the real flux \(\Phi\) controls loop interference in the physical-synthetic plane, whereas the imaginary gauge fields \(\gamma_x\) and \(\gamma_w\) control non-reciprocal envelopes and skin accumulation.

\begin{figure*}[t]
    \centering
    \maybegraphics{0.98\textwidth}{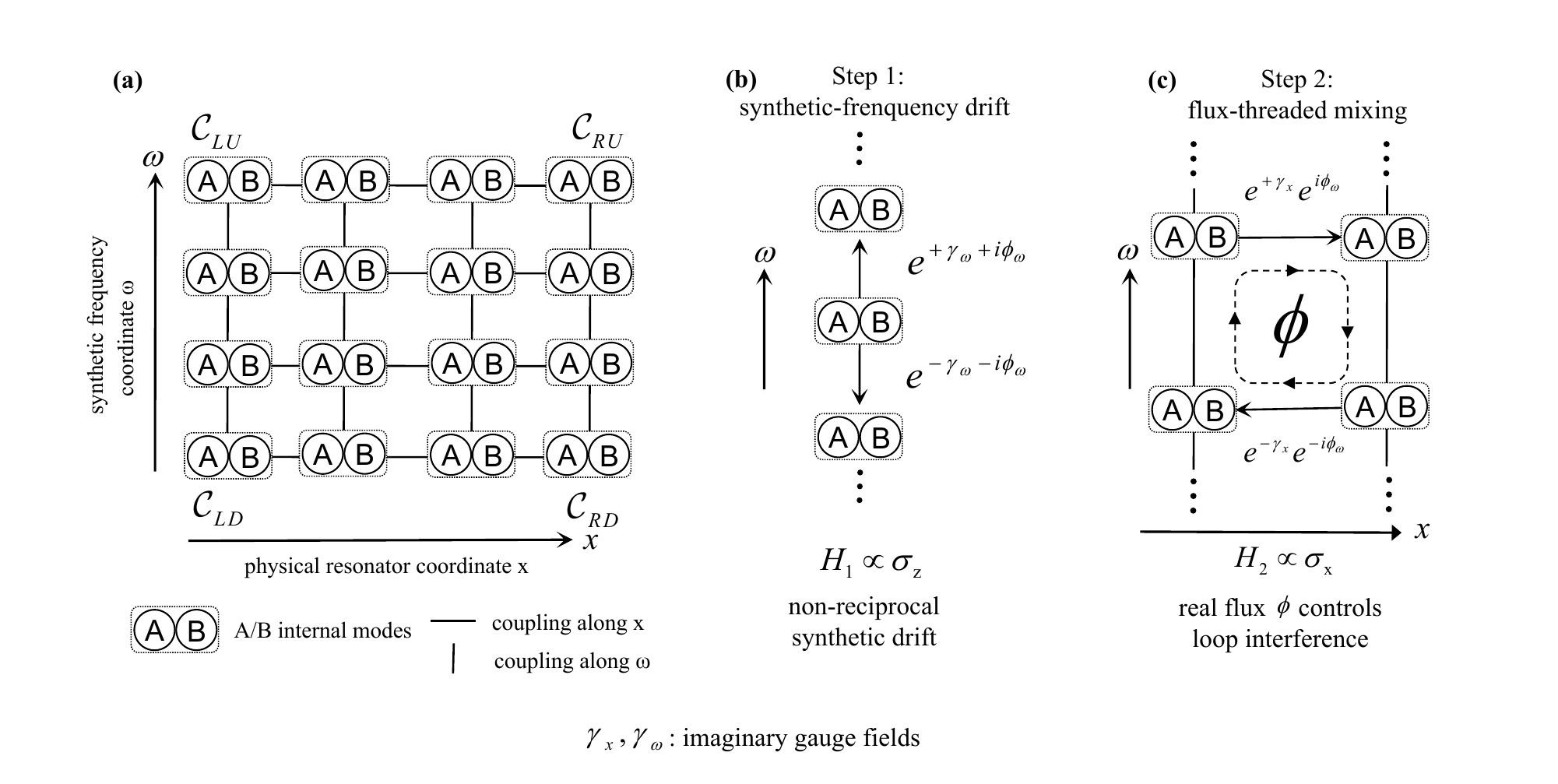}
    \vspace{-0.5cm}   
\caption{Concept of the photonic physical-synthetic Floquet lattice and the two-step complex-gauge drive.  Panel (a) shows the mode-space lattice.  The horizontal direction is the physical resonator coordinate \(x\), the vertical direction is the synthetic frequency coordinate \(w\), and each site contains two internal optical modes \(A\) and \(B\).  The four corner windows \(\calC_{\LD}\), \(\calC_{\LU}\), \(\calC_{\RD}\), and \(\calC_{\RU}\) are the local detection regions used for the doubled-period signal.  Panel (b) shows the first half step, which implements a non-reciprocal synthetic-frequency drift proportional to \(\sigma_z\); the upward and downward frequency-conversion amplitudes contain \(e^{+\gamma_w+\ii\phi_w}\) and \(e^{-\gamma_w-\ii\phi_w}\), respectively.  Panel (c) shows the second half step, which implements flux-threaded internal-mode mixing proportional to \(\sigma_x\); hopping along the physical direction carries the Landau-gauge phase \(e^{\ii\Phi w}\) and the non-reciprocal factor \(e^{\pm\gamma_x}\).  The real flux \(\Phi\) controls loop interference around a physical-synthetic plaquette, while \(\gamma_x\) and \(\gamma_w\) are imaginary gauge fields that control skin localization.}
    \label{fig1}
\end{figure*}

Both steps contain only \(\sigma_x\) and \(\sigma_z\) internal matrices.  With \(\Gamma=\sigma_y\), one has
\begin{equation}
    \Gamma H_j\Gamma=-H_j,
    \qquad j=1,2 .
\end{equation}
The symmetric time frames therefore satisfy a chiral Floquet relation.  This relation protects the special Floquet multipliers \(\lambda=+1\) and \(\lambda=-1\), corresponding to quasienergies \(0\) and \(\pi/T\), against symmetry-preserving perturbations.  It does not by itself determine the local optical visibility.  Figure~\ref{fig1} summarizes the architecture and fixes the notation used in the rest of the paper.  Panel (a) defines the physical-synthetic lattice and its four corner windows.  Panel (b) illustrates the synthetic-frequency drift step.  Panel (c) illustrates the flux-threaded mixing step.

\section{Corner modes and doubled-period optical readout}

The nonunitary Floquet eigenproblem is
\begin{equation}
    \UT|\psi_n^R\rangle=\lambda_n|\psi_n^R\rangle,
    \qquad
    \langle\psi_n^L|\UT=\lambda_n\langle\psi_n^L| .
    \label{eigen}
\end{equation}
A zero corner mode has \(\lambda_0\simeq+1\), while a \(\pi\) corner mode has \(\lambda_\pi\simeq-1\).  A local doubled-period response requires both modes to be coherently populated.  If
\begin{equation}
    |\Psi(0)\rangle
    =
    c_0|\psi_0^R\rangle
    +
    c_\pi e^{\ii\chi}|\psi_\pi^R\rangle,
\end{equation}
then stroboscopic evolution gives
\begin{equation}
    |\Psi(mT)\rangle
    \simeq
    c_0|\psi_0^R\rangle
    +
    (-1)^m c_\pi e^{\ii\chi}|\psi_\pi^R\rangle .
    \label{strobe_state}
\end{equation}
The relative sign in Eq.~\eqref{strobe_state} is the origin of the doubled-period optical response.

A local optical analyzer in a corner window \(\calC\) is
\begin{equation}
    O_{\calC}
    =
    \sum_{(x,w)\in\calC}
    |x,w\rangle\langle x,w|
    \otimes O_{\intm}.
    \label{analyzer}
\end{equation}
The choice of \(O_{\intm}\) specifies which internal-channel quadrature is read out.  For \(O_{\intm}=\sigma_x\), a local internal spinor \((a_A,a_B)^T\) gives
\begin{equation}
    (a_A^*,a_B^*)\sigma_x
    \begin{pmatrix}
        a_A\\
        a_B
    \end{pmatrix}
    =
    a_A^*a_B+a_B^*a_A
    =
    2\mathrm{Re}(a_A^*a_B).
    \label{sigmax_interference}
\end{equation}
Thus \(\sigma_x\) does not measure the total local intensity \(|a_A|^2+|a_B|^2\).  It measures the real part of the coherence between the two internal optical channels.  In an optical implementation this observable can be obtained by interfering the two outputs in the basis
\begin{equation}
    |+\rangle=\frac{|A\rangle+|B\rangle}{\sqrt2},
    \qquad
    |-\rangle=\frac{|A\rangle-|B\rangle}{\sqrt2}.
\end{equation}
If \(I_+\) and \(I_-\) are the two output intensities, then
\begin{equation}
    I_+-I_- = \langle \sigma_x\rangle .
\end{equation}
The operator \(O_{\calC}\) with \(O_{\intm}=\sigma_x\) is therefore a balanced interferometric analyzer restricted to the corner window \(\calC\).

The right-state corner visibility amplitude is
\begin{equation}
    S_{\calC}^{RR}
    =
    \langle\psi_0^R|O_{\calC}|\psi_\pi^R\rangle .
    \label{SRR}
\end{equation}
This amplitude is the local cross matrix element between the zero and \(\pi\) right eigenmodes.  It is the quantity that controls the doubled-period interference in the right-state optical field.  A biorthogonal amplitude
\begin{equation}
    S_{\calC}^{LR}
    =
    \langle\psi_0^L|O_{\calC}|\psi_\pi^R\rangle
\end{equation}
can also be defined.  The \(LR\) amplitude is useful as a biorthogonal response diagnostic, but it is not generally equivalent to \(S_{\calC}^{RR}\).  The right-right amplitude is the one directly tied to the right-state optical field distribution measured at output ports.

The raw balanced signal is
\begin{equation}
    I_{\calC}^{\mathrm{raw}}(m)
    =
    \langle\Psi(mT)|O_{\calC}|\Psi(mT)\rangle .
    \label{raw_signal}
\end{equation}
For the representative stroboscopic traces in Figs.~4 and 5, we use the normalized signal
\begin{equation}
    \bar I_{\calC}(m)
    =
    \frac{
    \langle\Psi(mT)|O_{\calC}|\Psi(mT)\rangle
    }{
    \langle\Psi(mT)|\Psi(mT)\rangle
    },
    \label{normalized_signal}
\end{equation}
so that a uniform overall loss or gain does not obscure the internal interference.  In the following visibility extraction, \(I_{\calC}(m)\) denotes the recorded signal used for the analysis; in the plotted traces this is \(\bar I_{\calC}(m)\).

The alternating component is extracted by a lock-in average with the reference sequence \((-1)^m\),
\begin{equation}
    \calA_{2T}^{RR}(\calC)
    =
    \left|
    \frac{1}{N_m}
    \sum_{m=m_0}^{m_0+N_m-1}
    (-1)^m I_{\calC}(m)
    \right|.
    \label{A2T}
\end{equation}
Indeed, if the local signal has the form
\begin{equation}
    I_{\calC}(m)
    =
    I_{\mathrm{dc}}+(-1)^m A_{\mathrm{ac}},
    \label{lockin_form}
\end{equation}
then
\begin{equation}
    \frac{1}{N_m}\sum_m(-1)^m I_{\calC}(m)
    \simeq
    A_{\mathrm{ac}}
    \label{lockin_result}
\end{equation}
for an even or sufficiently long sampling window.  Equation~\eqref{A2T} is therefore the stroboscopic Fourier component at quasifrequency \(\pi/T\).

Substituting Eq.~\eqref{strobe_state} into Eq.~\eqref{raw_signal}, the part alternating as \((-1)^m\) is
\begin{equation}
    A_{\mathrm{ac}}
    =
    2\mathrm{Re}
    \left[
    c_0^*c_\pi e^{\ii\chi}S_{\calC}^{RR}
    \right]
    \label{Aac_SRR}
\end{equation}
for a Hermitian analyzer \(O_{\calC}\).  For \(|c_0|=|c_\pi|=1/\sqrt2\), optimizing the relative phase \(\chi\) gives
\begin{equation}
    A_{\mathrm{ac}}^{\max}
    =
    |S_{\calC}^{RR}|.
    \label{Amax_SRR}
\end{equation}
Thus the local doubled-period signal is an interference measurement between the zero and \(\pi\) corner modes, not a measurement of a single corner-mode intensity.

\section{Non-Bloch higher-order invariant}

Under periodic boundaries, the Bloch factors are pure phases.  Under open boundaries with non-reciprocal conversion, the bulk ansatz must be continued to complex variables,
\begin{equation}
    \psi_{x,w}
    \propto
    \beta_x^x\beta_w^w u,
    \qquad
    \beta_\mu=r_\mu e^{\ii k_\mu},
    \qquad \mu=x,w .
    \label{nonbloch_ansatz}
\end{equation}
The closed curves traced by \(\beta_x\) and \(\beta_w\) are generalized Brillouin zones.  In the separable nearest-neighbor limit, the equal-modulus condition gives
\begin{equation}
    r_x=e^{-\gamma_x},
    \qquad
    r_w=e^{-\gamma_w} .
    \label{gbz_radii}
\end{equation}
These radii should be viewed as the simplest analytic limit.  In the flux-threaded physical-synthetic lattice, the real flux couples the two coordinates through the Landau-gauge phase in Eq.~\eqref{TxPhi}.  For a rational flux \(\Phi=2\pi p/q\), the magnetic unit cell contains \(q\) synthetic subcells and two internal modes, so the non-Bloch Floquet matrix is a \(2q\)-dimensional matrix in the magnetic cell.  In this general case the GBZ is obtained numerically from the characteristic equation together with the equal-modulus condition.  The resulting contours replace the ordinary unit circle in the topological windings below.

We use an edge non-Bloch construction for the higher-order index.  The construction has two layers.  The first layer determines whether a physical edge carries anomalous Floquet channels in the zero or \(\pi\) gap.  In the two symmetric time frames, the Floquet operators satisfy the chiral relation discussed in Appendix~B.  After rotating to the chiral basis, the off-diagonal blocks are denoted by \(q_1\) and \(q_2\).  Their non-Bloch windings along the physical GBZ are
\begin{equation}
    \widetilde W_\ell^x
    =
    \frac{1}{2\pi\ii}
    \oint_{\GBZ_x}
    \dd\log\det q_\ell(\beta_x),
    \qquad
    \ell=1,2 .
    \label{Wx}
\end{equation}
The two anomalous Floquet gap indices associated with the physical edge are then
\begin{equation}
    \widetilde\nu_0^x
    =
    \frac{\widetilde W_1^x+\widetilde W_2^x}{2},
    \qquad
    \widetilde\nu_\pi^x
    =
    \frac{\widetilde W_1^x-\widetilde W_2^x}{2} .
    \label{nux}
\end{equation}
Equations~\eqref{Wx} and \eqref{nux} are one-dimensional non-Bloch windings.  They do not assume that a two-dimensional Berry curvature is well defined in the non-Hermitian open-boundary problem.

The second layer asks whether the physical edge channel is itself topological along the synthetic direction.  We project the full Floquet drive onto the physical edge channel in the relevant quasienergy gap and evaluate a synthetic-direction edge winding, denoted by \(\widetilde\mu_\epsilon^w\), on the synthetic GBZ.  The resulting corner diagnostic is
\begin{equation}
    N_\epsilon^{\corner}(\Phi)
    =
    \widetilde\nu_\epsilon^x(\Phi)
    \widetilde\mu_\epsilon^w(\Phi),
    \qquad
    \epsilon=0,\pi .
    \label{Ncorner}
\end{equation}
This expression should be understood as an edge non-Bloch higher-order construction.  It is exact in the separable limit in which the two non-reciprocal directions can be treated independently, and it provides the bulk-corner diagnostic used below for the flux-threaded model with a rational magnetic unit cell.  Its prediction is the existence of zero or \(\pi\) corner modes.  It does not determine the right-eigenmode corner selected by skin accumulation, nor does it determine the local optical visibility.

\begin{figure*}[t]
    \centering
    \maybegraphics{0.92\textwidth}{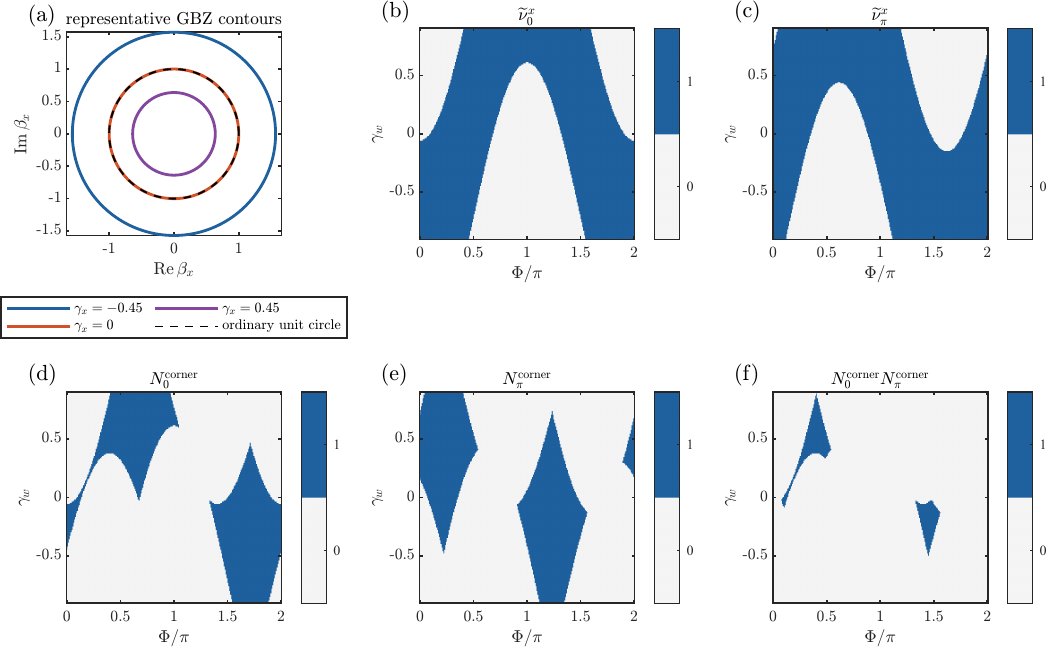}
    
\caption{Flux-resolved non-Bloch higher-order invariant.  Panel (a) shows representative generalized Brillouin-zone contours for different \(\gamma_x\), together with the ordinary unit circle as a dashed reference.  Panel (b) shows the first-layer physical-edge index \(\widetilde\nu_0^x\).  Panel (c) shows the corresponding index \(\widetilde\nu_\pi^x\).  Panel (d) shows the corner index \(N_0^{\corner}\) after including the synthetic-edge contribution.  Panel (e) shows \(N_\pi^{\corner}\).  Panel (f) shows the coexistence region \(N_0^{\corner}N_\pi^{\corner}=1\).  These maps are topological predictions obtained from non-Bloch cycles, not from finite open-boundary diagonalization.  Figure~\ref{fig3} tests these predictions by finite open-boundary spectra and shows the actual right-eigenmode corner selected by the imaginary gauge fields.}
    \label{fig2}
\end{figure*}

\begin{figure*}[t]
    \centering
    \maybegraphics{0.92\textwidth}{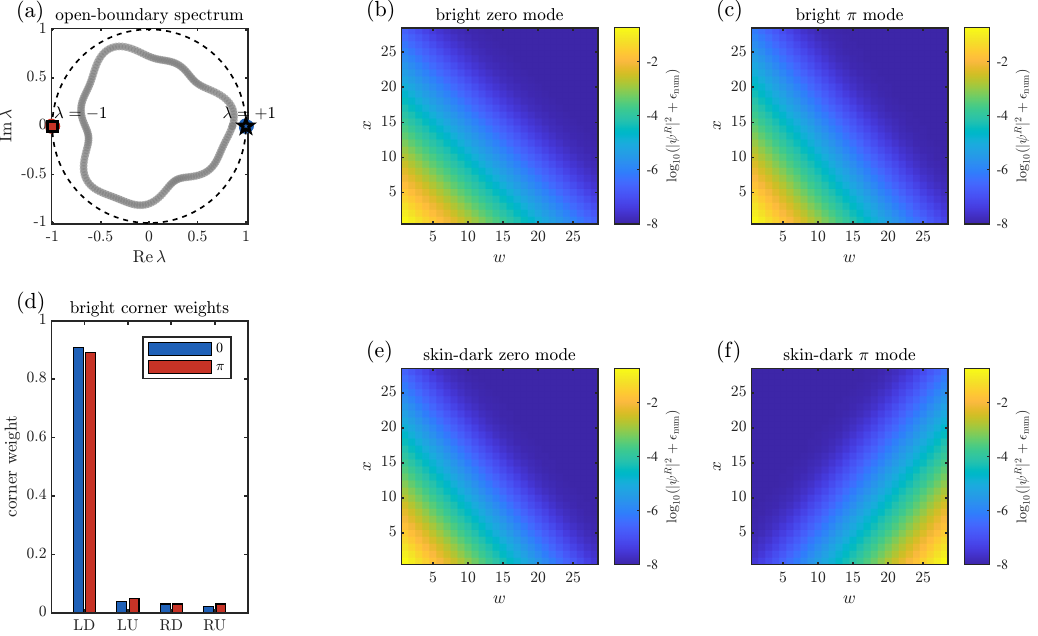}
    
\caption{Open-boundary spectrum and right-eigenmode profiles.  Panel (a) shows the finite open-boundary spectrum in the complex multiplier plane.  Gray points are bulk states.  Blue and red markers identify corner candidates near \(\lambda=+1\) and \(\lambda=-1\).  Panel (b) shows \(\log_{10}(|\psi_0^R(x,w)|^2+\epsilon_{\rm num})\) in the bright regime.  Panel (c) shows the corresponding \(\pi\) mode.  Here \(\epsilon_{\rm num}=10^{-8}\).  Panel (d) gives the four corner weights and confirms that the two modes occupy the same lower-left window.  Panel (e) shows the zero mode at a skin-dark point.  Panel (f) shows the \(\pi\) mode at the same point.  Both modes still exist, but the imaginary gauge fields select different synthetic corners for their right eigenmodes.  This spatial separation suppresses the local \(S_{\calC}^{RR}\) in any fixed corner window.}
    \label{fig3}
\end{figure*}

Figure~\ref{fig2} gives the topological prediction for the \(0/\pi\) corner-pair sector.  The logic of the figure follows Eq.~\eqref{Ncorner}: the non-Bloch cycles first determine the physical-edge Floquet indices, the projected edge problem then determines the synthetic-edge contribution, and their product predicts the corner sectors.  For stable numerical evaluation, the GBZ contour is sampled discretely, \(\det q_\ell\) is evaluated on the contour, and the phase of \(\det q_\ell\) is unwrapped continuously.  A winding number is not assigned when \(\det q_\ell\) approaches zero on the contour, since this signals a closing of the corresponding quasienergy gap.  The finite open-boundary test is shown in Fig.~\ref{fig3}.  There, eigenvalues near \(\lambda=+1\) and \(\lambda=-1\) identify zero and \(\pi\) candidates, while the corresponding right eigenvectors reveal which physical-synthetic corner is selected by the imaginary gauge fields.

\section{Topology versus optical visibility}

Equation~\eqref{Ncorner} predicts whether a zero or \(\pi\) corner mode exists.  It does not determine whether a local optical experiment sees a large doubled-period signal.  The reason is that the measured alternating signal is controlled by the local cross amplitude in Eq.~\eqref{SRR}, while topology fixes only the existence of the relevant corner eigenmodes.

To separate localization from visibility, we define the local right-state weight of a corner mode in a window \(\calC\) as
\begin{equation}
    W_{\calC}^{(\epsilon)}
    =
    \sum_{(x,w)\in\calC}
    |\psi_\epsilon^R(x,w)|^2,
    \qquad
    \epsilon=0,\pi .
    \label{local_weight}
\end{equation}
The weights \(W_{\calC}^{(0)}\) and \(W_{\calC}^{(\pi)}\) quantify whether the two modes are locally present in the same detection window.  They do not, by themselves, determine the interference visibility.  The latter is controlled by \(S_{\calC}^{RR}\), which also depends on the internal spinors and their flux-dependent relative phases.

We distinguish three regimes.  In a bright regime, the zero and \(\pi\) right modes both have order-one weights in the same corner window and the cross amplitude \(S_{\calC}^{RR}\) is large.  A local balanced analyzer then detects a large doubled-period response.

In a skin-dark regime, both modes exist but their right profiles are selected by different skin directions.  Suppose the zero mode is near \((1,1)\), while the \(\pi\) mode is near \((1,L_w)\).  In a fixed lower-left detection window, the \(\pi\) mode contributes only its exponentially small tail.  In this case
\begin{equation}
    W_{\calC_{\LD}}^{(0)}=O(1),
    \qquad
    W_{\calC_{\LD}}^{(\pi)}\sim e^{-2L_w/\xi_w},
\end{equation}
and the cross amplitude is suppressed as
\begin{equation}
    |S_{\calC_{\LD}}^{RR}|
    \propto
    e^{-L_w/\xi_w}.
    \label{skindark}
\end{equation}
The topological pair exists, but no fixed corner window contains both components with appreciable weight.

In a flux-dark regime, the two modes can remain co-localized while the local matrix element cancels by path interference.  The overlap has the schematic form
\begin{equation}
    S_{\calC}^{RR}(\Phi)
    =
    \sum_a S_a e^{\ii m_a\Phi},
    \label{pathsum}
\end{equation}
where \(m_a\) counts the oriented physical-synthetic plaquette area enclosed by path contribution \(a\).  At selected flux values, the terms in Eq.~\eqref{pathsum} can destructively interfere.  Then \(W_{\calC}^{(0)}\) and \(W_{\calC}^{(\pi)}\) can both be large while \(S_{\calC}^{RR}\simeq0\).  The corner modes remain present and locally co-localized, but the measured doubled-period interference amplitude becomes small.

\begin{figure*}[t]
    \centering
    \maybegraphics{0.94\textwidth}{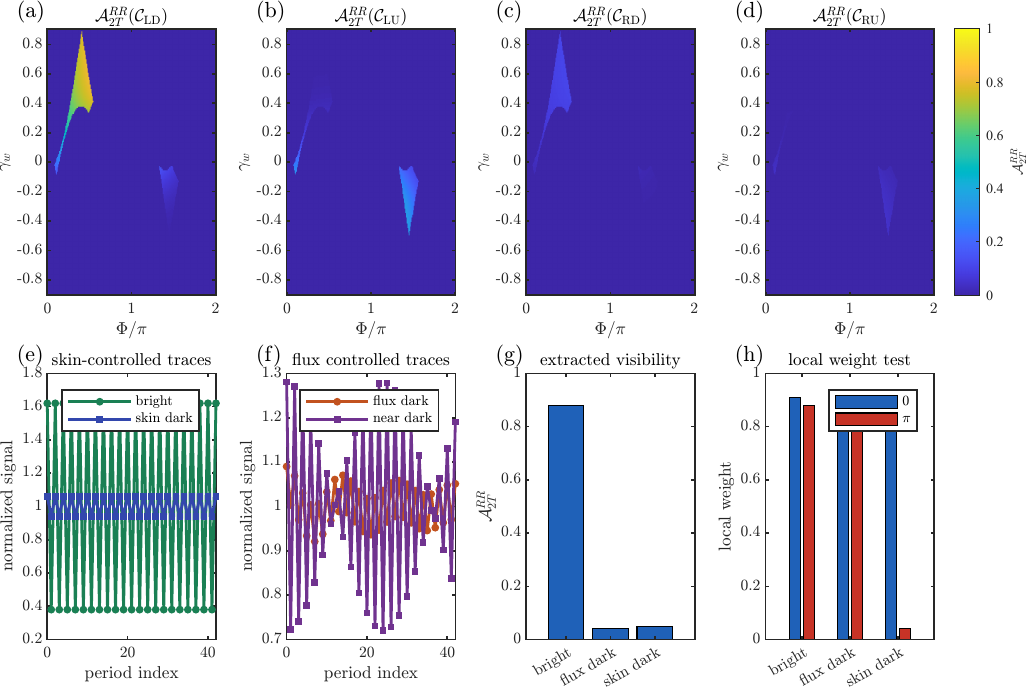}
    
\caption{Synthetic-flux and skin control of the doubled-period optical visibility.  Panels (a)--(d) show \(\calA_{2T}^{RR}(\calC)\) for \(\calC_{\LD}\), \(\calC_{\LU}\), \(\calC_{\RD}\), and \(\calC_{\RU}\), respectively, using a common color scale.  Panels (e) and (f) compare discrete stroboscopic traces in the bright, skin-dark, flux-dark, and near-dark regimes.  Panel (g) summarizes the extracted visibility at representative points.  Panel (h) compares the corresponding local zero and \(\pi\) weights.  Local weights are necessary but not sufficient for visibility.  Flux darkness comes from interference cancellation, whereas skin darkness comes from spatial separation.  Topological coexistence therefore does not imply local visibility.}
    \label{fig4}
\end{figure*}

Figure~\ref{fig4} is the main optical prediction.  Panels (a) through (d) show that the visible \(2T\) response can move between corner windows as the complex gauge is tuned.  Panels (e) through (h) then separate three data layers that should not be identified with one another.  These layers are the non-Bloch topological sector, the right-eigenmode corner location as measured by \(W_{\calC}^{(0,\pi)}\), and the measured interference visibility \(\calA_{2T}^{RR}\).  The bright regime has large local weights and large visibility.  The skin-dark regime has small visibility because one local weight is exponentially small.  The flux-dark regime has small visibility even though both local weights remain large, which shows that local co-localization is necessary but not sufficient for observing the doubled-period optical response.

\section{Defective doubled-period response}

The complex gauge also tunes an exceptional point in the corner dynamics.  It is useful to keep the roles of the one-period and two-period operators separate.  The one-period operator carries the anomalous Floquet sign structure,
\begin{equation}
    \UT|\psi_0^R\rangle\simeq +|\psi_0^R\rangle,
    \qquad
    \UT|\psi_\pi^R\rangle\simeq -|\psi_\pi^R\rangle .
    \label{one_period_signs}
\end{equation}
Therefore a zero mode and a \(\pi\) mode cannot directly coalesce as eigenvectors of \(\UT\) while preserving their quasienergy labels.  Their one-period multipliers are different.  The correct object for an exceptional point of the \(0/\pi\) corner subspace is the two-period propagator, because \((+1)^2\) and \((-1)^2\) are folded into the same multiplier sector.

We construct the effective two-period matrix by projecting \(\UT^2\) onto the corner subspace.  Let \(|\phi_0^R\rangle\) and \(|\phi_\pi^R\rangle\) be the two right corner states connected to the zero and \(\pi\) sectors, and let \(\langle\phi_0^L|\) and \(\langle\phi_\pi^L|\) be the corresponding left states.  We use the biorthogonal normalization
\begin{equation}
    \langle \phi_\alpha^L|\phi_\beta^R\rangle
    =
    \delta_{\alpha\beta},
    \qquad
    \alpha,\beta=0,\pi .
    \label{biorth_corner_norm}
\end{equation}
The projected propagator is
\begin{equation}
    \left[\UtwoT^{\eff}\right]_{\alpha\beta}
    =
    \langle\phi_\alpha^L|\UT^2|\phi_\beta^R\rangle,
    \qquad
    \alpha,\beta=0,\pi .
    \label{projected_U2T}
\end{equation}
This projection is meaningful when the two corner states are spectrally isolated from the remaining two-period spectrum.  This isolation is checked numerically by the open-boundary spectrum and the corner weights.

Any \(2\times2\) matrix can be decomposed into a scalar part and a traceless part.  We write
\begin{equation}
    \UtwoT^{\eff}
    =
    \Lambda_c I
    +
    K,
    \qquad
    K=
    \begin{pmatrix}
        \delta & g\\
        h & -\delta
    \end{pmatrix}.
    \label{K}
\end{equation}
Here \(\Lambda_c={\rm Tr}\,\UtwoT^{\eff}/2\) is the center of the two multipliers.  The parameter \(\delta\) is the relative two-period detuning of the two corner basis states, while \(g\) and \(h\) are the two directed couplings in the projected non-Hermitian corner subspace.  The real flux changes the phases of the effective coupling paths, whereas the imaginary gauge fields change the spatial overlap and the non-Hermitian asymmetry of the corner states.  In general \(g\neq h^*\).

The eigenvalues of \(K\) satisfy
\begin{equation}
    \det(K-\kappa I)=0,
    \qquad
    \kappa^2=\delta^2+gh .
\end{equation}
Thus the two eigenvalues of \(\UtwoT^{\eff}\) are
\begin{equation}
    \Lambda_\pm
    =
    \Lambda_c
    \pm
    \sqrt{\delta^2+gh},
    \label{Lambda_pm}
\end{equation}
so that
\begin{equation}
    \Lambda_+-\Lambda_-
    =
    2\sqrt{\delta^2+gh} .
    \label{splitting}
\end{equation}
We define the discriminant
\begin{equation}
    \Delta_{\rm EP}=\delta^2+gh .
    \label{discriminant}
\end{equation}
A second-order exceptional point occurs when
\begin{equation}
    \Delta_{\rm EP}=0,
    \qquad
    K\neq0 .
    \label{EP}
\end{equation}
The second condition is important.  If \(K=0\), then \(\UtwoT^{\eff}=\Lambda_c I\), which is an ordinary twofold degeneracy with two independent eigenvectors.  It is not an exceptional point.

The Jordan structure follows directly from
\begin{equation}
    K^2=(\delta^2+gh)I .
    \label{Ksquare}
\end{equation}
At the exceptional point, \(K^2=0\) but \(K\neq0\).  Hence the two-period propagator contains a nonzero nilpotent part,
\begin{equation}
    \begin{aligned}
        \UtwoT^{\eff}
        &=
        \Lambda_c(I+\calN),
        \qquad
        \calN=\Lambda_c^{-1}K,\\
        \calN^2&=0,
        \qquad
        \calN\neq0 .
    \end{aligned}
    \label{nilpotent_U2T}
\end{equation}
Repeated two-period evolution then gives
\begin{equation}
    \left(\UtwoT^{\eff}\right)^m
    =
    \Lambda_c^m(I+m\calN).
    \label{Jordan_power}
\end{equation}
The factor \(m\) is the dynamical signature of the defective corner subspace.  A generalized eigenvector is converted into the eigenvector by each two-period step.

When this two-period Jordan envelope is sampled at every period, it is multiplied by the one-period \(0/\pi\) sign in Eq.~\eqref{one_period_signs}.  The normalized or gain-corrected local signal therefore has the form
\begin{equation}
    I_{\calC}(n)
    =
    I_{\mathrm{dc}}(n)
    +
    (-1)^n
    \left[A_0+nA_J\right]
    +
    \cdots .
    \label{deftrace}
\end{equation}
The factor \((-1)^n\) records the anomalous \(0/\pi\) origin.  The algebraic envelope \(A_0+nA_J\) records the Jordan structure of the two-period propagator.  Thus the exceptional point does not create the doubled-period response.  It makes an existing \(0/\pi\) response defective.

Let \(|v_j^R\rangle\) and \(\langle v_j^L|\) be the right and left eigenvectors of \(\UtwoT^{\eff}\).  The eigenvector coalescence is diagnosed by the phase rigidity \cite{Heiss2012,Wiersig2014,Hodaei2017,Chen2017},
\begin{equation}
    r_j
    =
    \frac{
        |\langle v_j^L|v_j^R\rangle|
    }{
        \sqrt{
        \langle v_j^R|v_j^R\rangle
        \langle v_j^L|v_j^L\rangle}
    } .
    \label{phase_rigidity_main}
\end{equation}
At a second-order exceptional point, the two eigenvectors become self-orthogonal and \(r_j\to0\).  Uniform loss or gain in an optical implementation limits the useful observation window.  The defective envelope should therefore be extracted from transient stroboscopic traces before attenuation or amplification dominates the signal.

\begin{figure*}[t]
    \centering
    \maybegraphics{0.94\textwidth}{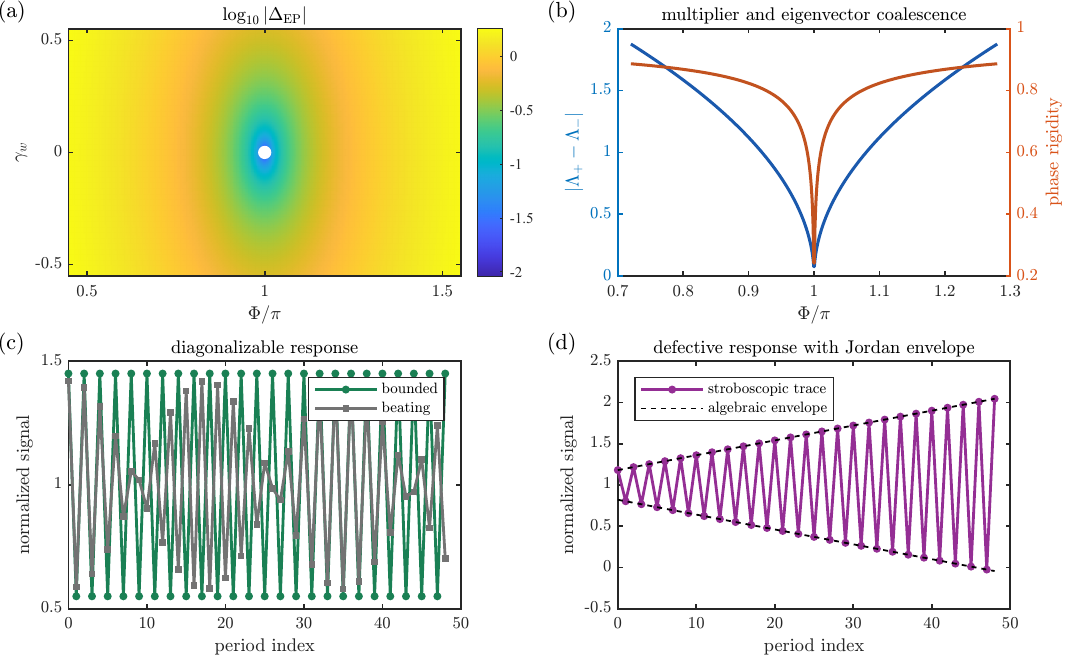}
    
\caption{Flux-tuned exceptional point and defective doubled-period response.  Panel (a) shows \(\log_{10}|\Delta_{\rm EP}|\), where \(\Delta_{\rm EP}=\delta^2+gh\).  The white marker identifies the exceptional point.  Panel (b) shows the two-period multiplier splitting and the phase rigidity across the exceptional point.  Their simultaneous collapse diagnoses eigenvalue and eigenvector coalescence.  Panel (c) compares bounded and beating responses away from the exceptional point.  Panel (d) shows the defective response and its algebraic envelope at the exceptional point.  The exceptional point is defined in the projected two-period corner propagator, not in the one-period \(0/\pi\) spectrum.  It does not create the \(2T\) signal.  It dresses an existing anomalous \(0/\pi\) response with a Jordan-block envelope.}
    \label{fig5}
\end{figure*}

Figure~\ref{fig5} completes the hierarchy.  The topology creates the corner pair, as in Fig.~\ref{fig2}.  The skin vector and flux determine whether that pair is locally visible, as in Figs.~\ref{fig3} and \ref{fig4}.  The exceptional point in the two-period corner propagator then determines whether the visible doubled-period response is diagonalizable or defective.

\section{Optical diagnostics}

All quantities in the main text can be extracted from the same numerical and experimental data.  The open-boundary spectrum gives the Floquet multipliers \(\lambda_n\).  The right corner weights identify where the emitted optical field accumulates.  The balanced local analyzer gives \(\calA_{2T}^{RR}(\calC)\).  The projected two-period matrix gives the exceptional-point discriminant and the phase rigidity.  The numerical workflow used for the figures follows this sequence: compute the flux-resolved non-Bloch corner diagnostic, identify the open-boundary corner profiles, evaluate the corner visibility amplitude, and project the two-period dynamics onto the corner subspace.

The required ingredients are available in dynamically modulated synthetic-frequency photonics \cite{Dutt2019,Dutt2020,Lustig2019,Dutt2020HOTI}.  The physical coordinate \(x\) is the resonator index and can be read out through spatial ports.  The synthetic coordinate \(w\) is the frequency-mode index and can be read out by frequency-resolved detection.  The two internal modes \(A\) and \(B\) may be implemented by two polarizations, two paths, clockwise and counterclockwise modes, or two coupled-resonator supermodes, depending on the platform.  The modulation frequency \(\Omega\) is chosen to match the resonator free spectral range or a selected near-resonant frequency spacing, so that the rotating-wave approximation leaves nearest-neighbor hopping along the synthetic frequency direction.

The two Floquet steps have direct optical interpretations.  The step proportional to \(\sigma_z\) requires internal-state-dependent frequency conversion, so that the two internal channels acquire opposite synthetic-frequency drifts.  The step proportional to \(\sigma_x\) requires coherent mixing between the internal modes.  The real synthetic flux \(\Phi\) is controlled by the phases of the modulation tones and appears as the Landau-gauge phase \(e^{\ii\Phi w}\) in the physical hopping.  The imaginary gauge fields \(\gamma_x\) and \(\gamma_w\) can be produced by asymmetric conversion amplitudes, gain/loss imbalance, or auxiliary lossy channels that make the effective forward and backward conversion strengths unequal \cite{ElGanainy2018,Miri2019}.  A finite synthetic boundary can be imposed by the finite bandwidth of the frequency comb, by spectral filtering, or by restricting the driven and detected frequency window.

The readout is also local in mode space.  Spatial ports resolve \(x\), a spectrum analyzer resolves \(w\), and an internal-mode interferometer realizes the \(O_{\intm}=\sigma_x\) analyzer by measuring the intensity difference in the \((A\pm B)/\sqrt2\) basis.  Hence the same optical measurement framework can access the topological sector, the skin-selected corner location, the flux-controlled doubled-period visibility, and the defective two-period dynamics.

\section{Conclusion}

We have studied a non-Hermitian physical-synthetic Floquet lattice in which a real synthetic flux and imaginary gauge fields act on the same anomalous \(0/\pi\) corner subspace.  The resulting boundary response is governed by a sequence of distinct mechanisms.  The edge non-Bloch construction determines whether zero and \(\pi\) corner modes exist under open boundaries.  The imaginary gauge fields select the physical-synthetic corner in which their right eigenmodes accumulate.  The real flux controls the local interference matrix element and therefore decides whether the doubled-period optical response is bright or dark.  The two-period exceptional point controls the diagonalizability of the corner dynamics and converts an ordinary \(0/\pi\) response into a defective doubled-period response with an algebraic envelope.

This separation is the main physical result.  A nonzero corner index is a statement about topological existence, not a guarantee of local optical visibility.  Large local weights are also not sufficient, since flux-induced destructive interference can suppress the measured \(2T\) component even when both corner modes occupy the same window.  Conversely, an exceptional point does not generate the subharmonic sign alternation.  It modifies the two-period envelope of an already present anomalous Floquet response.  These distinctions give a practical route for using photonic synthetic dimensions to test non-Bloch topology, skin-selected localization, flux-controlled interferometry, and defective dynamics within one measurement framework.

\begin{acknowledgments}
This work was supported by the National Natural Science Foundation of China (Grant No. 12275193).  The authors thank the members of the quantum physics group at Tianjin Normal University for helpful discussions.

\end{acknowledgments}

\section*{Data Availability}
The numerical data and codes that support the findings of this work are available from the corresponding author upon reasonable request.

\appendix

\section{Coupled-mode origin of the synthetic-frequency hopping}

This appendix explains why the dynamical variables in the synthetic-frequency lattice are the modal amplitudes rather than the spatial mode profiles.  The unmodulated resonator has fixed eigenmodes \(\bm f_w(\bm r)\) with eigenfrequencies \(\omega_w\).  The electric field is expanded as
\begin{equation}
    \bm E(\bm r,t)
    =
    \sum_w c_w(t)\bm f_w(\bm r)e^{-\ii\omega_w t}
    +\mathrm{c.c.},
    \label{app_field_expansion}
\end{equation}
where \(c_w(t)\) is the slowly varying envelope in the rotating frame.  The functions \(\bm f_w(\bm r)\) define the fixed optical-mode basis.  The coefficients \(c_w(t)\) are the coordinates of the field in that basis and are the dynamical variables measured through the output sideband powers.

For a weak time-dependent dielectric perturbation \(\delta\varepsilon(\bm r,t)\), projection of Maxwell's equation onto the unmodulated modes gives the rotating-frame coupled-mode equation
\begin{equation}
    \ii\dot c_m(t)
    =
    \sum_n
    V_{mn}(t)e^{\ii(\omega_m-\omega_n)t}c_n(t),
    \label{app_cmt_general}
\end{equation}
with the overlap matrix element
\begin{equation}
    V_{mn}(t)
    =
    -\frac{\omega_0}{2}
    \int d^3\bm r\,
    \bm f_m^*(\bm r)\cdot
    \delta\varepsilon(\bm r,t)
    \bm f_n(\bm r),
    \label{app_overlap}
\end{equation}
where the overall sign is a phase convention.  Equation~\eqref{app_cmt_general} has the form of a tight-binding Schrödinger equation in mode space.  A basis state \(|w\rangle\) is an optical frequency mode, and the amplitude \(c_w(t)\) is the wave-function amplitude on that mode-space site.

For an equally spaced frequency ladder \(\omega_{w+1}-\omega_w=\Omega\), a modulation at frequency \(\Omega\),
\begin{equation}
    \delta\varepsilon(t)
    =
    \delta\varepsilon^{(+)}e^{-\ii(\Omega t-\varphi)}
    +
    \delta\varepsilon^{(-)}e^{+\ii(\Omega t-\varphi)},
\end{equation}
selects nearest-neighbor hopping along the synthetic frequency direction.  The factor \(e^{-\ii\Omega t}\) cancels the rotating-frame phase for the process \(w\to w+1\), while \(e^{+\ii\Omega t}\) cancels the phase for the reverse process \(w+1\to w\).  Rapidly oscillating nonresonant terms are neglected in the rotating-wave approximation.  The retained terms give
\begin{equation}
    H_w
    =
    \sum_w
    \left[
    J_w^+|w+1\rangle\langle w|
    +
    J_w^-|w\rangle\langle w+1|
    \right].
    \label{app_Hw}
\end{equation}
The modulation phase becomes the Peierls phase of the synthetic hopping, while the modulation amplitude sets the hopping strength.  Reciprocal, lossless conversion gives \(J_w^-=(J_w^+)^*\).  Non-reciprocal conversion gives \(|J_w^+|\neq |J_w^-|\), which is parameterized as \(J_w^\pm=J_w e^{\pm\gamma_w\pm\ii\phi_w}\) and realizes the imaginary gauge field used in Eq.~\eqref{H1}.

\section{Floquet algebra and chiral time frames}

For two steps that anticommute with \(\Gamma=\sigma_y\), the symmetric time frames are
\begin{equation}
    U_1
    =
    e^{-\ii H_1T/4}
    e^{-\ii H_2T/2}
    e^{-\ii H_1T/4},
\end{equation}
and
\begin{equation}
    U_2
    =
    e^{-\ii H_2T/4}
    e^{-\ii H_1T/2}
    e^{-\ii H_2T/4}.
\end{equation}
They obey
\begin{equation}
    \Gamma U_\ell\Gamma=U_\ell^{-1},
    \qquad
    \ell=1,2 .
\end{equation}
After rotating to a chiral basis, each \(U_\ell\) has an off-diagonal block \(q_\ell\).  The windings of \(\det q_\ell\) determine the two anomalous Floquet gaps through Eq.~\eqref{nux}.

\section{Non-Bloch numerical winding}

For a rational flux \(\Phi=2\pi p/q\), the Landau-gauge phase enlarges the magnetic unit cell.  The unit cell contains \(q\) synthetic subcells and two internal modes, so each symmetric-frame Floquet matrix becomes a \(2q\)-dimensional non-Bloch matrix.  We denote it by \(U_\ell(\beta_x,\beta_w)\), where \(\ell=1,2\) labels the symmetric time frame and \(\beta_\mu=r_\mu e^{\ii k_\mu}\) is the non-Bloch variable in direction \(\mu=x,w\).

The numerical winding used in Eq.~\eqref{Wx} is computed as follows.  First, one determines or approximates the relevant GBZ contour.  In the separable nearest-neighbor limit we use the analytic radii in Eq.~\eqref{gbz_radii}.  In the general flux-threaded case one obtains the contour from the characteristic equation and the equal-modulus condition for the dominant roots.  Second, the contour is sampled as a sequence of points \(\beta_j\), and the complex numbers
\begin{equation}
    z_{\ell,j}=\det q_\ell(\beta_j)
\end{equation}
are evaluated.  Third, the phase \(\arg z_{\ell,j}\) is unwrapped continuously along the contour.  The winding is then
\begin{equation}
    \widetilde W_\ell
    =
    \frac{1}{2\pi}
    \Delta \arg \det q_\ell .
    \label{app_winding_discrete}
\end{equation}
If \(\min_j |z_{\ell,j}|\) falls below the numerical tolerance, the winding is not assigned.  This situation means that \(\det q_\ell\) touches the origin on the contour and the corresponding quasienergy gap is closing.  The integer value of \(\widetilde W_\ell\) is only meaningful away from such gap-closing regions.

The projected synthetic-edge winding \(\widetilde\mu_\epsilon^w\) is obtained in the same way after projecting the full drive onto the physical edge channel in gap \(\epsilon=0,\pi\).  Operationally, one constructs the edge subspace from a strip geometry that is open along the physical direction and periodic or non-Bloch along the synthetic direction, forms the projected edge Floquet operator, rotates it to the chiral basis when needed, and evaluates the phase winding of its off-diagonal block along \(\GBZ_w\).

As a consistency check, the predicted pair \((N_0^{\corner},N_\pi^{\corner})\) is compared with finite open-boundary diagonalization.  A finite-system eigenstate is counted as a zero or \(\pi\) corner candidate when its multiplier lies within a small window around \(+1\) or \(-1\) and its corner weight exceeds a fixed threshold.  Repeating this comparison for several system sizes distinguishes robust corner sectors from finite-size leakage.  Mismatches are expected only near gap-closing lines or when the localization length becomes comparable to the system size.  This comparison is the finite-size validation underlying the use of Eq.~\eqref{Ncorner} as the corner diagnostic in the main text.

\section{Visibility and exceptional-point diagnostics}

The distinction between bright, skin-dark, and flux-dark regimes is diagnosed by comparing three quantities.  The first is the topological pair \((N_0^{\corner},N_\pi^{\corner})\).  The second is the pair of right-state local weights \(W_{\calC}^{(0)}\) and \(W_{\calC}^{(\pi)}\).  The third is the cross amplitude \(S_{\calC}^{RR}\).  Bright behavior requires all three ingredients: a topological pair, co-localization of the two right eigenmodes, and a large local interference matrix element.  Skin-dark behavior has small \(S_{\calC}^{RR}\) because one mode has exponentially small local weight.  Flux-dark behavior has small \(S_{\calC}^{RR}\) even though both weights are large, because the flux-dependent terms in the local overlap cancel.

The exceptional point is diagnosed by the projected two-period corner propagator defined in Eq.~\eqref{projected_U2T}.  The discriminant \(\Delta_{\rm EP}=\delta^2+gh\) gives the multiplier splitting through Eq.~\eqref{splitting}.  At the exceptional point, \(\Delta_{\rm EP}=0\) and \(K\neq0\), so \(K\) is nilpotent rather than zero.  The phase rigidity in Eq.~\eqref{phase_rigidity_main} diagnoses the accompanying eigenvector coalescence.  The time-domain signature is the defective doubled-period trace in Eq.~\eqref{deftrace}.

\section{Projection of the two-period corner propagator}
\label{app:projection_U2T}

This appendix summarizes the numerical construction of \(\UtwoT^{\eff}\).  For a fixed set of parameters, the finite open-boundary Floquet operator \(\UT\) is diagonalized together with its left eigenvectors.  Two right eigenvectors are selected by their multipliers near \(+1\) and \(-1\) and by their large corner weights.  These states define \(|\phi_0^R\rangle\) and \(|\phi_\pi^R\rangle\).  The corresponding left eigenvectors are rescaled so that
\begin{equation}
    \langle\phi_\alpha^L|\phi_\beta^R\rangle=\delta_{\alpha\beta},
    \qquad
    \alpha,\beta=0,\pi .
\end{equation}
The projected two-period matrix is then
\begin{equation}
    \UtwoT^{\eff}
    =
    \begin{pmatrix}
    \langle\phi_0^L|\UT^2|\phi_0^R\rangle &
    \langle\phi_0^L|\UT^2|\phi_\pi^R\rangle \\
    \langle\phi_\pi^L|\UT^2|\phi_0^R\rangle &
    \langle\phi_\pi^L|\UT^2|\phi_\pi^R\rangle
    \end{pmatrix} .
\end{equation}
From this matrix one obtains
\begin{align}
    \Lambda_c
    &=
    \frac{1}{2}\mathrm{Tr}\,\UtwoT^{\eff},\\
    \delta
    &=
    \frac{1}{2}
    \left[
    (\UtwoT^{\eff})_{00}
    -
    (\UtwoT^{\eff})_{\pi\pi}
    \right],\\
    g&=(\UtwoT^{\eff})_{0\pi},
    \qquad
    h=(\UtwoT^{\eff})_{\pi0} .
\end{align}
The discriminant \(\Delta_{\rm EP}=\delta^2+gh\), the multiplier splitting, and the phase rigidity are then evaluated from this \(2\times2\) matrix.  This construction assumes that the selected two-dimensional corner sector is separated from the remaining two-period spectrum.  Close to a bulk or edge gap closing, the projection is no longer a reliable isolated-corner description.

\section{Visibility formula and interferometric readout}

This appendix gives the elementary derivation of the visibility extraction used in Sec.~III.  For a local spinor \(|a\rangle=(a_A,a_B)^T\), the internal analyzer \(\sigma_x\) gives
\begin{equation}
    \langle a|\sigma_x|a\rangle
    =
    a_A^*a_B+a_B^*a_A
    =
    2\mathrm{Re}(a_A^*a_B).
\end{equation}
It therefore measures the real part of the internal-channel coherence.  In the interferometric basis \(|\pm\rangle=(|A\rangle\pm|B\rangle)/\sqrt2\), the two output amplitudes are \(a_\pm=(a_A\pm a_B)/\sqrt2\).  Hence
\begin{equation}
    |a_+|^2-|a_-|^2
    =
    a_A^*a_B+a_B^*a_A
    =
    \langle a|\sigma_x|a\rangle .
\end{equation}
The corner analyzer \(O_{\calC}\) is thus implemented by summing this balanced intensity difference over the output channels belonging to the corner window \(\calC\).

For a coherent \(0/\pi\) corner superposition, Eq.~\eqref{strobe_state} gives the signal
\begin{align}
    I_{\calC}(m)
    &=
    I_{\mathrm{dc}}
    +
    (-1)^m
    \left[
    c_0^*c_\pi e^{\ii\chi}S_{\calC}^{RR}
    +
    c_\pi^*c_0 e^{-\ii\chi}(S_{\calC}^{RR})^*
    \right]
    \\
    &=
    I_{\mathrm{dc}}
    +
    (-1)^m A_{\mathrm{ac}},
\end{align}
where
\begin{equation}
    A_{\mathrm{ac}}
    =
    2\mathrm{Re}
    \left[
    c_0^*c_\pi e^{\ii\chi}S_{\calC}^{RR}
    \right].
\end{equation}
Multiplying by the reference sequence \((-1)^m\) converts the alternating term into a constant.  Thus
\begin{equation}
    \frac{1}{N_m}\sum_{m=m_0}^{m_0+N_m-1}
    (-1)^m I_{\calC}(m)
    \simeq
    A_{\mathrm{ac}},
\end{equation}
when the sampling window contains many periods or an integer number of doubled periods.  This is the lock-in extraction of the stroboscopic component at quasifrequency \(\pi/T\).

If \(|c_0|=|c_\pi|=1/\sqrt2\), one may choose the relative phase \(\chi\) to cancel the phase of \(c_0^*c_\pi S_{\calC}^{RR}\).  The largest alternating amplitude is then
\begin{equation}
    A_{\mathrm{ac}}^{\max}=|S_{\calC}^{RR}|.
\end{equation}
This shows why the right-state cross amplitude, rather than the separate local weights alone, is the relevant optical visibility measure.

\end{document}